\documentclass[fleqn,10pt]{wlscirep}
\usepackage[utf8]{inputenc}
\usepackage[T1]{fontenc}

\usepackage{graphicx}
\usepackage{epstopdf}
\usepackage{subfigure}
\usepackage{rotating}
\usepackage{amsmath}
\usepackage{amssymb}
\usepackage{xcolor}

\newcommand{\matN}{\mathcal{N}}
\newcommand{\matW}{\mathcal{W}}

\newcommand{\fp}[1]{{\color{black}#1}}

\begin{document}

\title{The role of vascular complexity on optimal junction exponents}

\author{Jonathan Keelan}

\author{James P. Hague}
\affil{School of Physical Science, The Open University, MK7 6AA, UK}

\affil[*]{Jim.Hague@open.ac.uk}

\begin{abstract}
We examine the role of complexity on arterial tree structures, determining globally optimal vessel arrangements using the Simulated AnneaLing Vascular Optimization (SALVO) algorithm, which we have previously used to reproduce features of cardiac and cerebral vasculatures. Fundamental biophysical understanding of complex vascular structure has applications to modelling of cardiovascular diseases, and for improved representations of vasculatures in large artificial tissues. In order to progress in-silico methods for growing arterial networks, we need to understand the stability of computational arterial growth algorithms to complexity, variations in physiological parameters such as tissue demand, and underlying assumptions regarding the value of junction exponents. We determine the globally optimal structure of two-dimensional arterial trees; analysing sensitivity of tree morphology and optimal bifurcation exponent to physiological parameters. We find that, for physiologically relevant simulation parameters, arterial structure is stable, whereas optimal junction exponents vary. We conclude that the full complexity of arterial trees is essential for determining the fundamental properties of vasculatures. These results are important for establishing that optimisation-based arterial growth algorithms are stable against uncertainties in physiological parameters, while identifying that optimal bifurcation exponents (a key parameter for many arterial growth algorithms) are sensitive to complexity and the boundary conditions dictated by organs.
\end{abstract}

\maketitle

\section{Introduction}

\fp{Vascular systems are highly complex, multiscale systems, with dominant physics that changes with length scale.} Within a typical organ, vascular trees connect major arteries of $\sim10$mm diameter to huge numbers of tiny arterioles with width of $\sim 10-100\mu$m. The physics of large vessels is often dominated by pulsatile flows and turbulence, while small vessels are microfluidic \cite{Nakamura2014}. Evolution must account for this complex and multiscale physics when optimizing arterial networks, since efficient vasculatures are crucial for supplying oxygen to tissues.

\fp{Computational techniques and analytic expressions for describing these complex and multiscale networks have applications in physiology, tissue engineering, and medical diagnosis.} Beyond the desire to understand the fundamental biological properties of vascular networks, deviations from optimal flow conditions could be a sign of underlying disease, and vascularization is a key issue limiting the growth of large engineered tissues.  

\fp{The challenge is that the complex and multiscale structures of vascular networks are difficult to reproduce {\it in-silico}.} Within organs, there can be hundreds of thousands of arterioles dependent on every major artery. The number of possible combinations of vessels associated with these connections is enormous. Since it is not possible to deterministically search all these combinations for all but the smallest trees, stochastic methods are needed. We previously introduced the SALVO algorithm to find the globally optimal structure of arteries using simulated annealing to overcome these problems \cite{keelan2016, keelan2019}.


\fp{Vascular networks are primarily constructed from bifurcations \cite{bifexp}, which can be characterized by defining a bifurcation exponent (also known as radius exponent and junction exponent), $\gamma$.} The radii of the two output vessels, $r_{\rm out,A}$ and $r_{\rm out,B}$, are related to the radius of the input vessel, $r_{\rm in}$, via,
\begin{equation}
r_{\rm in}^{\gamma} = r_{\rm out,A}^{\gamma}+r_{\rm out,B}^{\gamma}.
\label{eqn:murraylaw}
\end{equation}

\fp{Murray carried out a single-vessel analysis, which when combined with conservation of flow, shows that when flow in vessels is approximately laminar (Poiseuille flow) the optimal junction exponent, $\gamma_{\rm opt}=3$ \cite{Murray1926a}.} In Murray's analysis, there are two competing contributions to the metabolic demand of vessels: the power dissipated during Poiseuille flow, and the metabolic cost of maintaining a volume of blood. The former is minimized for wide vessels and the latter for narrow vessels, so the actual radius is a compromise.

%
%

\fp{In living systems, the bifurcation exponent is often measured to deviate from three \cite{Nakamura2014}, which is not fully understood, although several factors are known to modify $\gamma_{\rm opt}$ in single-vessel analyses.} Inclusion of pulsatile flow, elastic wall vessels, and turbulence contribute to reduction of the optimal junction exponent to $\gamma_{\rm opt}=2.33$ \cite{Nakamura2014}. A key assumption leading to allometric scaling laws is that cross-sectional area is conserved, i.e. $\gamma_{\rm opt}=2$ \cite{west1997general}. 


\fp{Values of $\gamma$, measured in many vascular networks, are larger than expected from single-segment analyses.}  In some organs, $\gamma$ is found to be slightly greater than three \cite{Nakamura2014,keelan2019}. To our knowledge, no explanation of this effect is currently available, since corrections to flow in single artery analyses to include turbulence, pulsatile flow, and elastic wall vessels, lead to $\gamma_{\rm opt}<3$. This suggests a role for complexity in vascular network analysis.

\fp{We propose that, in order to fully understand the optimal branching exponents in vascular trees, it is essential to take into account the complexity of the entire arterial network in an organ, and the boundary conditions imposed by the organism.} A single vessel is part of a much larger arterial tree for an organ, that is in turn part of an organism, and the role of this additional complexity is poorly understood. The metabolic demand of the organ determines the blood flow to the organ. The radius of the primary artery supplying that organ is determined by a compromise between the whole organism and the organ. These two properties define boundary conditions for arterial growth algorithms.

\fp{In this paper we carry out a theoretical and numerical analysis of the optimal bifurcation exponent for large and complex arterial trees with physiologically measured boundary conditions.} The work goes beyond previous analyses \cite{Murray1926a,Zamir1979,Wischgoll2009,Zamir2001,Horsfield1989,Nakamura2014}, by optimizing entire trees, rather than a single arterial bifurcation. The constraints on flow and radius of root vessels in real organs are also taken into account. Further to this analysis, we use the SALVO algorithm \cite{keelan2016,keelan2019} to determine the numerically exact globally optimal bifurcation exponent.

\section{Methods}

\subsection{Power cost}

The arterial tree is divided into straight segments and bifurcations, with Poiseuille flow assumed within
segments. The power cost for a single arterial tree segment
experiencing Poiseuille flow is,
\begin{equation}
  W_{j} = m_b \pi r_{j}^2 l_{j} + \frac{8 \mu f_{j}^{2} l_{j}}{\pi r_{j}^{4}}
  \label{eq:singlesegment}
\end{equation}
where $j$ denotes a segment, $r_{j}$ the segment radius, $l_{j}$
its length, $f_{j}$ its volumetric flow, $m_b$ the metabolic power
cost of blood, and $\mu$ the dynamic viscosity of blood. The power cost associated with bifurcations is
neglected.

The total cost, $\matW$,
of an arterial tree is the sum of these individual segment costs,
\begin{equation}
\matW = \sum_{j \in \{\rm segments\} } W_{j}.\label{eq:costfunc}
\end{equation}

\subsection{Murray's law}

Murray's law is derived by optimizing total cost in a single segment (Eq. \ref{eq:singlesegment}). By differentiating with respect to $r_{j}$,
\begin{equation}
    \frac{\partial W_{j}}{\partial r_{j}}=2m_{b}\pi f_{j}^{2}l_{j}-\frac{32\mu f_{j}^2l_{j}}{\pi r_{j}^{5}}.
\end{equation}
When $\partial W_{j}/\partial r_{j}=0$, the optimal $r_{j}$ can be found. This leads to a relation for flow in terms of $r_{j}$
\begin{equation}
f_{j}=\frac{m_{b}^{1/2}\pi}{4\mu^{1/2}}r_{j}^{3} 
\end{equation}

In the following analysis, we will assume that $l=l_{\rm root}r^{\alpha}/r_{\rm root}^{\alpha}$, where $l_{\rm root}$ and $r_{\rm root}$ are the length and radius of the root segment respectively, and $\alpha$ is the length--radius exponent. This slightly modifies the preceding argument, so that,
\begin{equation}
f_{j}=\frac{m_{b}^{1/2}\pi}{2(2\mu)^{1/2}}r_{j}^{3}\sqrt{\frac{2+\alpha}{4-\alpha}} = \frac{f_{\rm root}}{r_{\rm root}^{3}}{r_{j}^{3}},
\end{equation}
where $f_{\rm root}$ is the flow through the root segment.

\subsection{SALVO}

The power of the numerical SALVO algorithm is that the generated trees are globally optimized and therefore represent the lowest possible power cost, allowing the effects of evolutionary optimization to be investigated. While the globally optimized solution represents an idealized evolutionary endpoint, insight into the compromise associated with optimizing the competing costs of the different metabolic requirements associated with maintaining a complicated vasculature will be gained by this analysis.

The  (SALVO) algorithm developed in earlier papers
\cite{keelan2016,keelan2019} can be used to generate arterial trees in
the 2D plane. In this section, an outline of this algorithm is given. The algorithm is similar to the
approach for growing cardiac and cerebral vasculature\cite{keelan2016,keelan2019}, with some differences relating to the use of fixed nodes to supply tissue.

Equation \ref{eq:costfunc} is the cost function at the core of the SALVO algorithm. In this paper we study an idealized two-dimensional (2D) piece of `tissue'. Leaf nodes are fixed in place, and there is no supply penalty. Similarly, there is no penalty for tissue penetration of large vessels, since all vessels lie within the 2D tissue. The root node of the tree is fixed to the corner of a square region of side $a$. In contrast to earlier use of the algorithm, a Poisson disc process is used to place leaf nodes \cite{bridson2007a}. The whole 2D region
was accessible by nodes, with only metabolic- and flow-related
penalties consistent with Eq. \ref{eq:singlesegment}.

\begin{figure}
\centering
  \includegraphics[width=60mm]{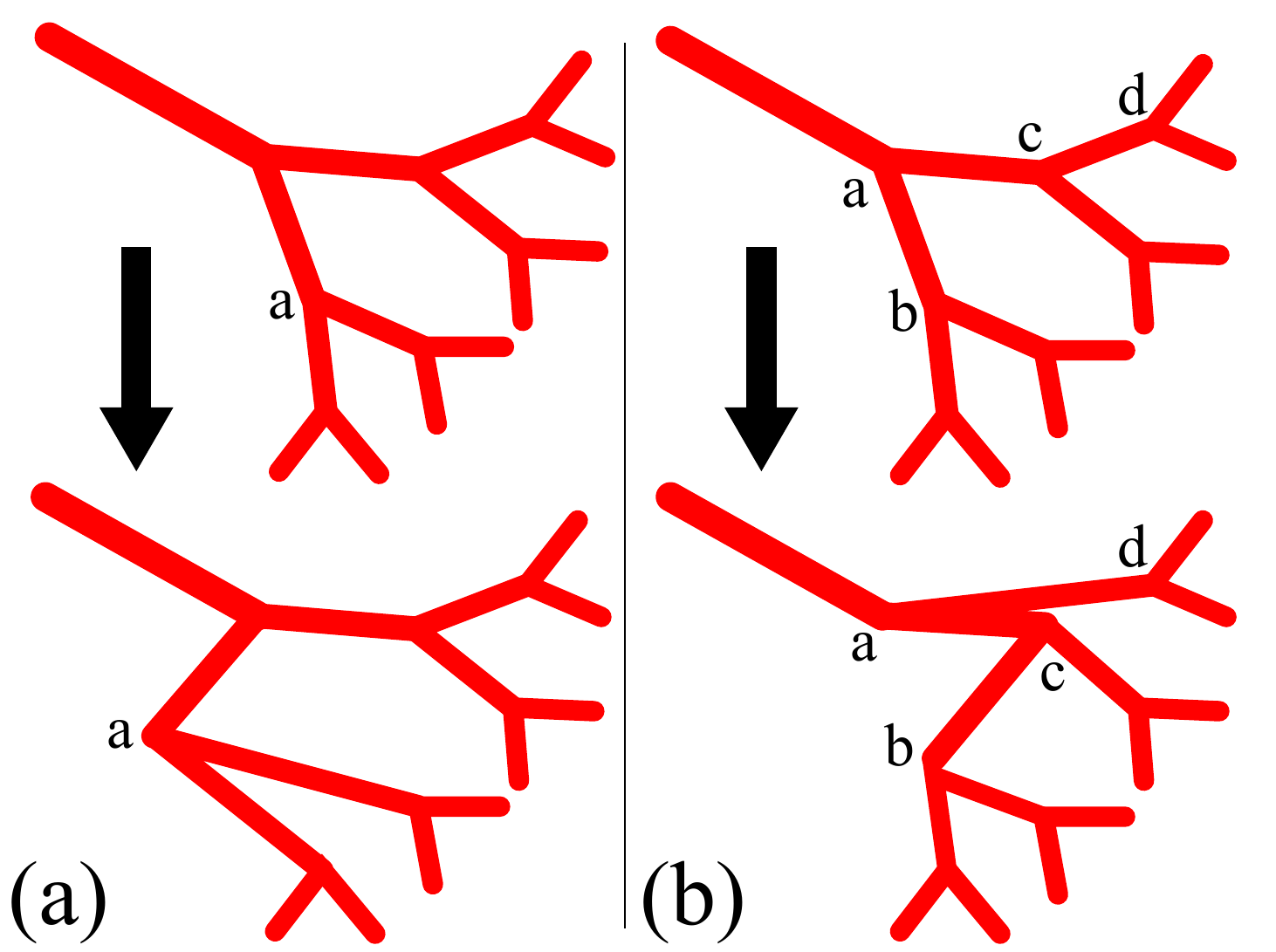}
\caption{Two types of update, that move node coordinates and swap the parent segments of nodes respectively, are required for ergodicity. The figure shows a summary of these updates. In panel (a) node a is moved. In panel (b) parents of two nodes (nodes b and d) are swapped. The parent of node b is node a, and the parent of node d is node c. After the swap, the parent of node b is node c and the parent of node d is node a.}
\label{fig:updatessummary}
\end{figure}

On each iteration, modifications to the binary tree are attempted by either (1) moving a node or (2) changing the tree structure by swapping node connections. These updates are sufficient to ensure ergodicity. Updates are summarized in Fig.
\ref{fig:updatessummary} and Table
\ref{table:sweepparameters}. The root node is never updated. In this version of the algorithm, leaf nodes are never moved. We use simulated annealing to optimize the cost function \cite{siman}. 

Acceptance of the updates is determined according to the probability,
\begin{equation}
P_{\theta,\theta+1} = {\rm min}\left\{\exp(\frac{-\Delta \matW^{(\theta,\theta+1})}{T}),1\right\}\label{annealing}
\end{equation}
where $\Delta \matW^{(\theta,\theta+1)}=\matW^{(\theta+1)}-\matW^{(\theta)}$ is the change in cost associated with modifying the tree from configuration $(\theta)$ to
configuration $(\theta+1)$, respectively. $T$ is the annealing temperature,
which is slowly reduced using the common exponential schedule, $T_{\theta + 1} = \epsilon T_{\theta}$ where $\theta$ is the iteration number, $\epsilon = \exp( \ln T_{0} - \ln T_{\Theta})/\Theta$, $\Theta$ the total number of iterations, and $T_{0}$ ($T_{\Theta}$) are the initial (final) temperatures.

SALVO was implemented in C++ making full use of the 2011 standard library (g++ version 7.4.0 compiled with the -O3 flag). Generated trees were analyzed using Python 3.7. A Threadripper 2990WX processor was used for the calculations, with calculations for different $\gamma$, $\Omega$ and tree sizes sent to different threads by a Python script. The optimization of a 500 node tree takes approximately 30 minutes, and a 5000 node tree takes approximately 11 hours on a single thread for $10^8$ updates.

\section{Analytical results}

\subsection{Formalism and simplifications}
  
Arteries can be grouped together, so that each group comprises
arteries with identical properties (e.g length, diameter, flow). In a
real arterial system, this would not be true, but it would still be
possible to group arteries with similar lengths, radii, and flows
together.
  
Without loss of generality,
the total power can be rewritten as,
\begin{equation}
\matW = \sum_{j \in \{r,l,q\}} \matN(r_j,l_j,f_j) \left( m_b \pi r_{j}^{2} l_j + \frac{8 \mu f_{j}^{2} l_j}{\pi r_{j}^{4}} \right), \label{eq:costequation}
\end{equation}
where $\matN(r_j,l_j,f_j)$ is the number of arterial segments with identical
radii, lengths and flows.

Under the
restriction that the flow in all leaf nodes is identical and equal to
$f_{\rm leaf}$, the flow in each segment is,
\begin{equation}
  f_{n} = n f_{\rm leaf}
\end{equation}
where $n$ is an integer and represents the total number of leaf nodes
downstream of the segment.

Comparing Eq. \ref{eqn:murraylaw} with flow conservation, a radius--flow relation is identified:
\begin{equation}
  f_{n} = f_{\rm leaf} (r_{n}/r_{\rm leaf})^{\gamma}
  \label{eqn:radiusflow}
\end{equation}
thus,
\begin{equation}
  r_{n}=r_{\rm leaf}\left(f_{n}/f_{\rm leaf}\right)^{1/\gamma} = r_{\rm leaf}n^{1/\gamma}.
  \label{eqn:flowradius}
\end{equation}

Experimental data suggest that the length of an arterial segment is
proportional to a power of the radius,
\begin{equation}
  l_{n} = l_{\rm leaf} \left(r_{n}/r_{\rm leaf}\right)^{\alpha} = l_{\rm leaf} \left(f_{n}/f_{\rm leaf}\right)^{\alpha/\gamma},
  \label{eqn:lengthscaling}
\end{equation}
where the value of the exponent $\alpha$ is typically close to 1.0 \cite{Kamiya2007,Nakamura2014}. 

By substituting Eqs. \ref{eqn:lengthscaling}, the power required to
maintain blood flow through a segment depends only on the flow $f_{n}$,
\begin{eqnarray}
  W_{n} = W(f_{n}) & = & m_{b}\pi r_{\rm leaf}^{2}l_{\rm leaf} \left(f_{n}/f_{\rm leaf}\right)^{(2+\alpha)/\gamma}\nonumber\\
  & & +\frac{8\mu l_{\rm leaf}}{\pi r_{\rm leaf}^{4}} f_{n}^{2} \left(f_{n}/f_{\rm leaf}\right)^{(\alpha-4)/\gamma}.
\end{eqnarray}

Thus, the dimensionless \emph{metabolic ratio}, defined as $\Omega = m_{b}\pi^2 r_{\rm leaf}^{6}/8\mu f_{\rm leaf}^2$, along with $N$, controls location in parameter space. 
\begin{eqnarray}
W_{n} & = & C \left(\Omega n^{(2+\alpha)/\gamma}+n^{2+(\alpha-4)/\gamma}\right)\\
& = & C n^{1+(\alpha-1)/\gamma}\left(\Omega n^{3/\gamma-1}+n^{1-3/\gamma}\right)
\label{eqn:seesymmetryclearly}
\end{eqnarray}
$C=\frac{8 \mu  f_{\rm leaf}^2 l_{\rm leaf}}{\pi r_{\rm leaf}^4}$. Both $C$ and $\Omega$ are defined in terms of the leaf node.

A similar ratio for the root node, $\Omega_{\rm root}=m_{b}\pi^2 r_{\rm root}^{6}/8\mu f_{\rm root}^2$ can be defined for convenient contact with experiment. The values $r_{\rm root}$ and $f_{\rm root}$ are often known from experiment, e.g. Doppler ultrasound, and $N$ can be estimated. This ratio can be related to $\Omega$ via $\Omega_{\rm root}=N^{6/\gamma-2} \Omega$.

The total power requirement is,
\begin{equation}
  \matW = \sum_{n} \matN_n W_{n}.
  \label{eqn:totalpowergeneral}
\end{equation}
$\matN_{n}$ is the number of segments with flow $n f_{\rm leaf}$, and
simplifies the function $\matN(r_i,l_i,f_i)$. For any tree structure,
$N$ is always the number of leaf nodes, so $\matN_{1}=N$. There
is always a single root node with total flow $N f_{\rm leaf}$, so $\matN_{N} = 1$. The remaining $\matN_n$ are dependent on the structure of the tree. At each bifurcation, flow conservation requires that $n_{\rm in}=n_{\rm out,1} + n_{\rm out,2}$, so $n$ is an integer.

Total power is linear in length scale, so the structure of the power landscape (including the location of any minima with respect to $\gamma$) is independent of $a$. It is the global minimum with respect to $\gamma$ that sets the structure of the tree, and when locating the minimum, $\partial\matW/\partial\gamma=0$, so the factors of $l_{\rm leaf}$ simply cancel, thus making the solution independent of $a$. Changes in $r_{\rm leaf}$ can be absorbed into the ratio $m_b/\mu$ and thus are similar to changing the metabolic requirements of the organ \cite{keelan2016}.

There are two special cases: fully symmetric and fully asymmetric. In the first case, identified as a fully
symmetric tree, the flow is split evenly at each
bifurcation. For the case which we shall identify as fully asymmetric, a single leaf node emerges at each bifurcation and the rest of the flow passes down the other bifurcation. We will explore these special cases in the following two sections.

\subsection{Fully symmetric vascular tree}

In a fully symmetric tree, all of the segments with flow $n$ exist at the same bifurcation layer. Each layer, $m$, has $2^{m}$ segments, where $m$ is the number of bifurcations upstream of that layer ($m = 0$ at the root segment). Within a layer,
all segments have the same flow, and thus the same radius and length. The tree has a total of $M$ layers.

The total power cost can be determined by substituting the definitions $\matN_{n} =2^{m}$ if $n=2^{M-m}$ otherwise $\matN_{n}=0$, into Eq. \ref{eqn:totalpowergeneral},
\begin{equation}
  \matW = C\sum_{m = 0}^{M} 2^{m} \left( \Omega 2^{(M-m)(2+\alpha)/\gamma}+2^{(M-m)(2+(\alpha-4)/\gamma)} \right). \label{eq:costreducedconstgam}
\end{equation}

Thus, by summing the geometric series, the total power cost for a fully symmetric tree is,
\begin{equation}
\matW = 2^{M} C\left(\Omega\frac{2^{-M(1-(2+\alpha)/\gamma)}-1}{1-2^{1-(2+\alpha)/\gamma}}+ \frac{2^{M(1+(\alpha-4)/\gamma)}-1}{1-2^{-(1+(\alpha-4)/\gamma)}}\right)
\label{eqn:symmetriccost}
\end{equation}

\subsection{Fully asymmetric tree}

The total power cost of the fully asymmetric tree may be calculated by noting that each discrete flow is represented once for all $n$, so
$\matN_{n} = 1$, except there are $N$ leaf nodes so $\matN_{1}=N$.

Substitution into Eq. \ref{eqn:totalpowergeneral} gives,
\begin{equation}
  \matW = C\left(\sum_{n=1}^{N} (\Omega n^{(2+\alpha)/\gamma}+n^{2+(\alpha-4)/\gamma})
  + (\Omega+1)(N-1)\right)
\end{equation}

So the total power cost for an asymmetric tree is
\begin{eqnarray}
    \matW & = & C\left(\Omega H_{N}^{(-(2+\alpha)/\gamma)}+H_{N}^{(-(2+(\alpha-4)/\gamma))}\right) \nonumber\\
    & & \hspace{20mm} + C(\Omega+1)(N-1),
    \label{eqn:asummetriccost}
\end{eqnarray}
where $H_{n}^{(r)}$ is the generalized harmonic function, $\sum_{k=1}^{n} 1/k^{r}$.
%

\subsection{Optimal bifurcation exponent}

The optimal value of $\gamma$ is obtained by solving $\partial \matW/\partial \gamma=0$ for Eqs. \ref{eqn:symmetriccost} and \ref{eqn:asummetriccost}.
 We numerically solve $\partial \matW/\partial \gamma=0$ using Mathematica's contour plot routines (Mathematica v8.0.4.0, Wolfram). Computations are much faster for the symmetric than the asymmetric trees. 
 
 \begin{figure}
    \centering
    \includegraphics[width=80mm]{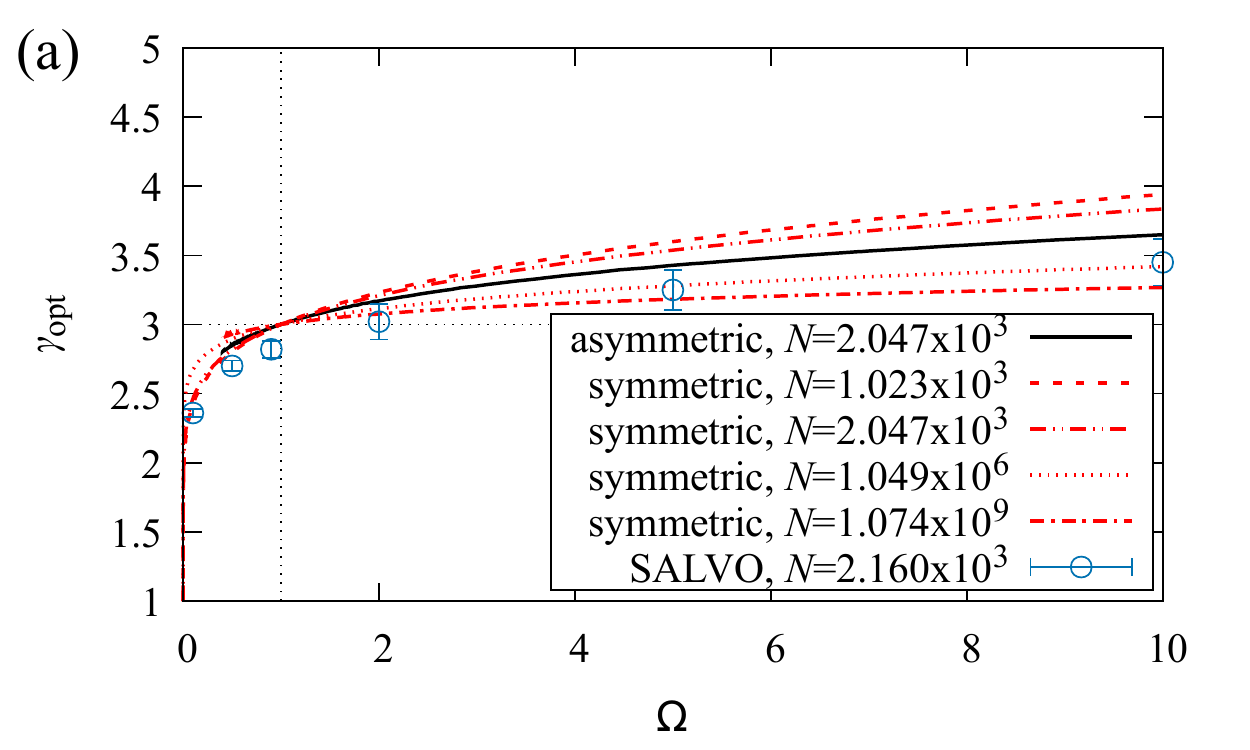}
        \includegraphics[width=80mm]{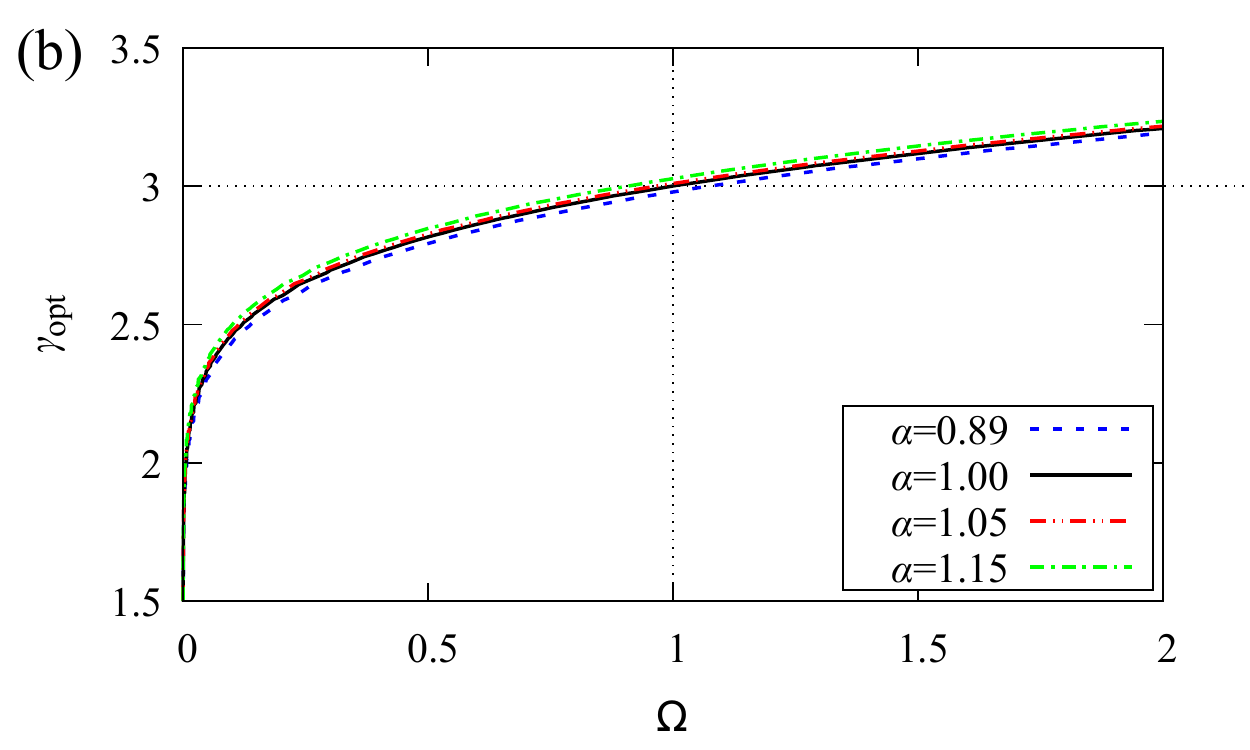}
        \caption{(a) Deviations from Murray's law ($\gamma_{\rm opt}=3$) depend strongly on changes in the metabolic ratio, but are insensitive to the structure of the tree. The figure shows a comparison of $\gamma_{\rm opt}$ vs $\Omega$ for fully symmetric, asymmetric and numerical trees. (b) Optimal bifurcation exponent is insensitive to changes in $\alpha$.}
    \label{fig:symmetricgamma}
        \label{fig:gammaalpha}
\end{figure}
 
 The optimal bifurcation exponent is strongly dependent on the metabolic ratio, $\Omega$, which can change due to physiological boundary conditions on flow and radius at the input vessels. These constraints may be due to limits in the size of the largest vessel and changing flow demands of tissue.  Figure \ref{fig:symmetricgamma}(a) shows the optimal value of $\gamma$. When $\Omega=1$ and $\alpha=1$ the result of Murray's law ($\gamma_{\rm opt}=3$) is recovered. 
 
 $\gamma_{\rm opt}$ is qualitatively unchanged by the structure of the tree. Results for asymmetric and symmetric trees with $N=2.047\times 10^{3}$ follow essentially the same functional forms. The optimal bifurcation exponent for the asymmetric tree is closer to $\gamma=3$ than the symmetric tree. Also shown in Fig. \ref{fig:symmetricgamma}(a) are numerical values from SALVO, which will be discussed later.

The optimal bifurcation exponent is modified up or down from $\gamma=3$ by changes in the length exponent, $\alpha$ (Fig. \ref{fig:gammaalpha}(b)). This structural effect potentially has implications for the value of $\gamma_{\rm opt}$ in organs, since $\alpha$ can vary with organ type, with estimates ranging from $0.89-1.15$. In practice, changes in $\gamma_{\rm opt}$ for this variation in $\alpha$ are far less than the error for measurements of $\gamma$ and changes in $\alpha$ can essentially be neglected.

\begin{figure}
    \centering
    \includegraphics[width=80mm]{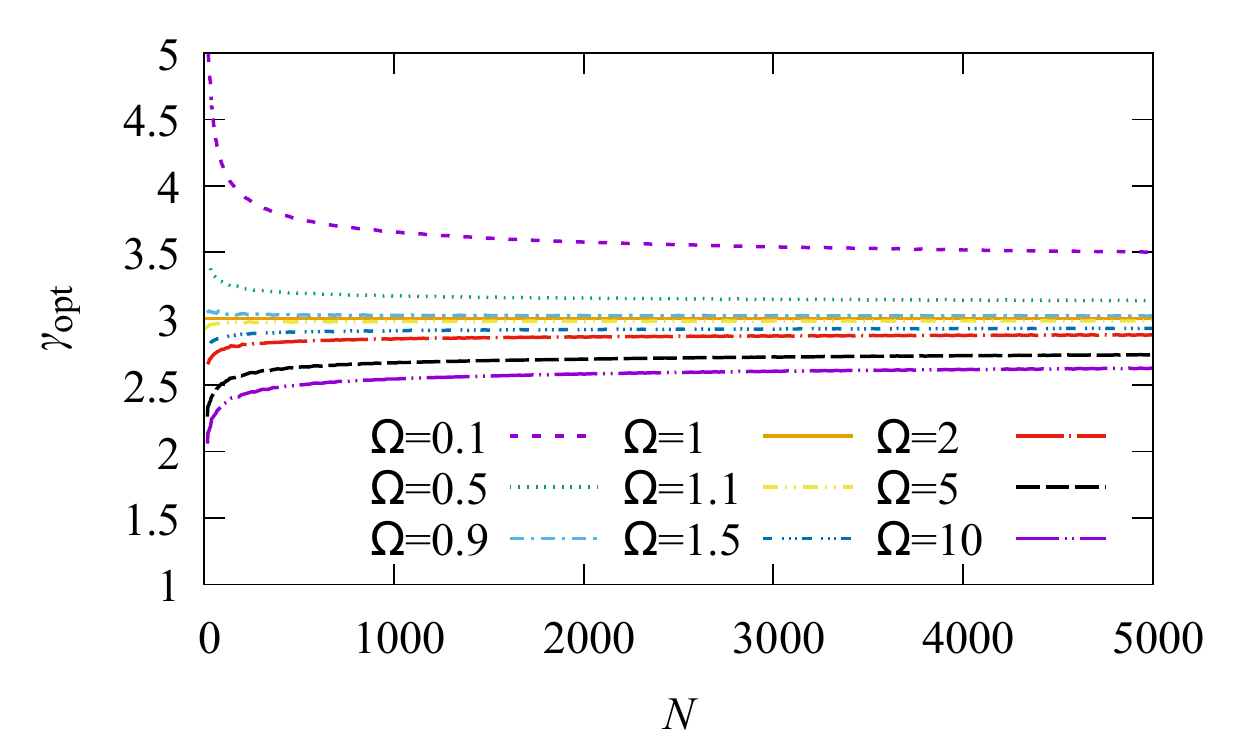}
    \caption{Deviations from Murray's law are largest for small trees and strongly dependent on changes in the metabolic ratio. The figure shows $\gamma_{\rm opt}$ vs $N$ for a fully asymmetric tree.}
    \label{fig:asymmetricgammavsN}
\end{figure}

Deviations from Murray's law are larger for smaller trees and strongly dependent on changes in the metabolic ratio. The larger the tree, the closer to Murray's law $\gamma_{\rm opt}$ becomes. Figure \ref{fig:asymmetricgammavsN} shows variation of $\gamma_{\rm opt}$ with $N$ for fully symmetric trees. For vascular tree sizes of between $10^{3}$ and $10^{6}$ segments, which are typical in organs, $\gamma_{\rm opt}$ ranges between 2 and 4. 

\section{Numerical results}

The generation of globally optimal trees using a numerical algorithm helps to test analytic expressions, and provides additional morphological measures that can be used to understand arterial networks. In this section, we use SALVO to investigate the role of vascular complexity and physiological boundary conditions on the properties of globally optimal trees. Several properties of the numerically generated trees are investigated. We determine the sensitivity of globally optimal trees to $\Omega$ and $\gamma$. Through examination of $\matW_{\rm tot}$, we compute $\gamma_{\rm opt}$ for complex trees. For each value of $\gamma$ and $\Omega$ investigated, arterial trees with up to 5000 nodes were generated. Table \ref{table:sweepparameters} summarizes the parameters used for the numerical calculations.

\begin{table}
  \centering
  \caption{Simulation parameters and their ranges.}
  \label{table:sweepparameters}
  \begin{tabular}{|c | c | c |}
 \hline
 Name & symbol & range \\
\hline
 bifurcation exponent & $\gamma$ & 1.0-5.0\\
 metabolic ratio & $\Omega$ & 0.1-10\\ 
number of leaf nodes & $N$ & 100-5000 \\
blood viscosity & $\mu$ & $3.6\times 10^{-3}$ Pa s\\
tissue size & $a$ & 1cm\\
    SA steps & $\Theta$ & $10^{8}$ ($10^{9}$ for checks) \\
    SA initial `temperature' & $T_{0}$ & 1 Js$^{-1}$\\
    SA final `temperature' & $T_{\Theta}$ & $10^{-12}$ Js$^{-1}$ \\
    short move distance & $d_{\mathrm{move}}$ & $0.05$mm \\
    long move distance & $d_{\mathrm{move}}$ & $0.5$mm \\
    short move node weight &  & 0.3 \\
    long move update weight &  & 0.2 \\    
    swap update weight &  & 0.5 \\    \hline
  \end{tabular}
\end{table}

\begin{figure}
    \centering
    \includegraphics[width=120mm]{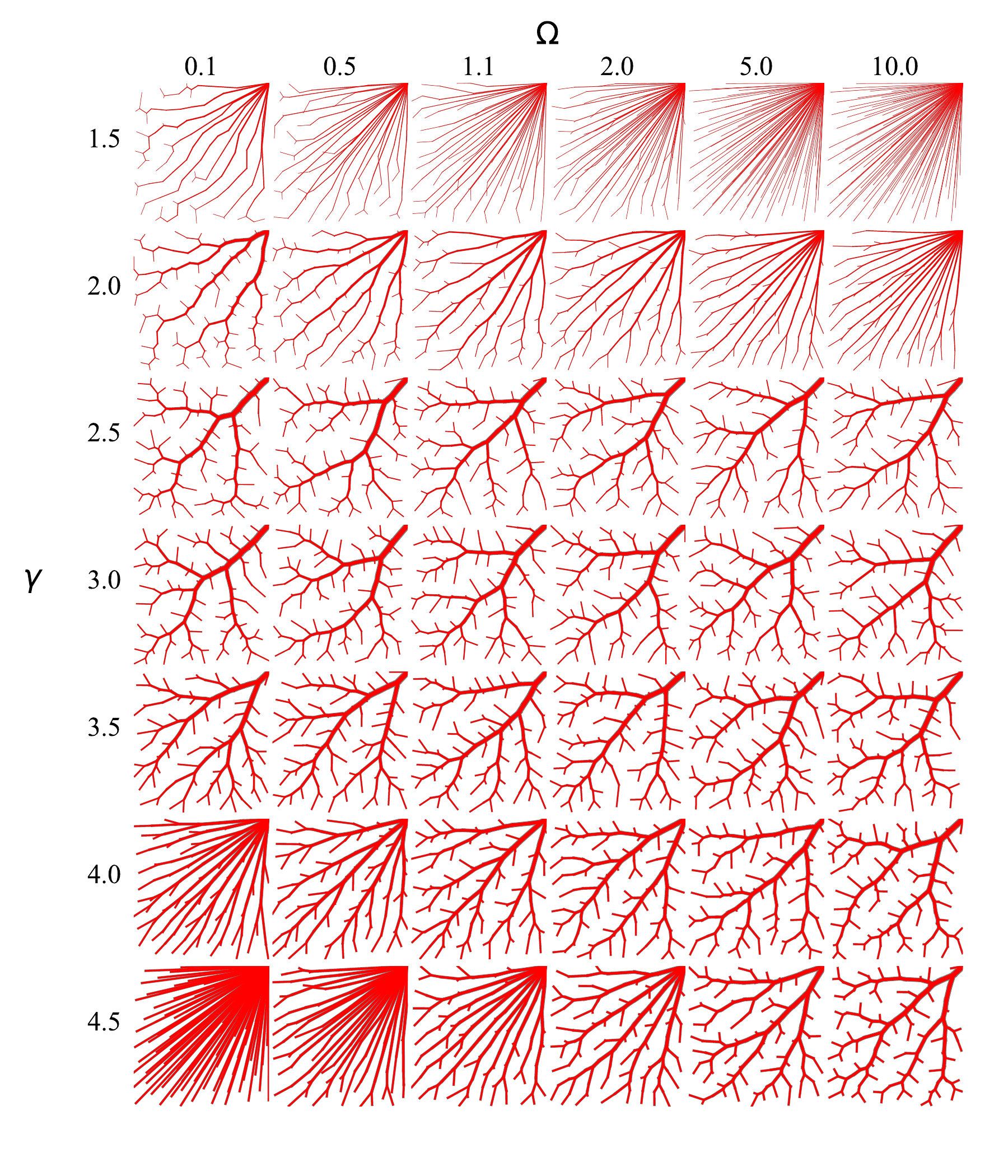}
    \caption{The structure of the globally optimal vasculature varies with $\gamma$ and $\Omega$. Trees have size $N=100$. Radii are normalized by the root radius for easier visualization.}
    \label{fig:varyntrees}
\end{figure}

Three sectors of the parameter space have qualitatively different tree structures (Fig. \ref{fig:varyntrees}): (1) for $\gamma\lesssim 2, \Omega>1$ and $\gamma\gtrsim 4, \Omega<1$, long and narrow leaf nodes originate from the vicinity of the root node; (2) for $\gamma\lesssim 2, \Omega<1$ and $\gamma\gtrsim 4, \Omega>1$ asymmetric and tortuous branches dominate; (3) for $2\lesssim \gamma \lesssim 4$ trees have a branching structure similar to the kinds of vasculature seen in living tissue. In the figure, the vessel widths are normalized to the root radius to improve visibility.

Trees with $\Omega\gg 1$, $\gamma<2$ and $\Omega\ll 1$, $\gamma>4$ are very similar, which is not a coincidence, and can be explained by examining the structure of Eq. \ref{eqn:seesymmetryclearly}. When $\alpha\approx 1$, the power in a segment is $W_{n} = C n(\Omega n^{3/\gamma-1}+n^{1-3/\gamma})$. For $\gamma>3$, the exponents (which involve $3/\gamma-1$) have opposite sign to those for $\gamma<3$. So after the substitutions $\Omega=1/\Omega'$, $\gamma=3\gamma'/(2\gamma'-3), C'=\Omega C$, $W_{n} = C' n(\Omega' n^{3/\gamma'-1}+n^{1-3/\gamma'})$, and the sum has an equivalent structure. The substitution is determined by identifying where $1-3/\gamma=3/\gamma'-1$. Since the prefactor $C'$ scales the entire sum, then the minima of $\matW$ and thus the results for $\gamma,\Omega$ and $\gamma',\Omega'$ are identical. This symmetry is only approximate if $\alpha\neq 1$.

For $\gamma\lesssim 2$ and small $\Omega$ (and $\gamma\gtrsim 4$ and large $\Omega$), the tree structure is highly asymmetric, with long trunks snaking through leaf node sites (see top left panels in Fig. \ref{fig:varyntrees}). This is due to the domination of the $n^{1-3/\gamma}$ term due to Poiseuille flow for low $\gamma$, and the $n^{1-3/\gamma'}$ metabolic cost term for large $\gamma$. Thus terms with large $n$ (i.e. thick trunks) are favored.

For $\gamma\lesssim 2$ and large $\Omega$ (and $\gamma\gtrsim 4$ and small $\Omega$), long leaf segments connect root and leaf nodes (see top right panels in Fig. \ref{fig:varyntrees}). This is due to the domination of the $n^{3/\gamma-1}$ term due to metabolic maintenance of blood for low $\gamma$, and the $n^{3/\gamma'-1}$ Poiseuille term for large $\gamma$. The term related to metabolic maintenance of blood (with $n^{3/\gamma-1}$) dominates. Thus, terms with small $n$ (i.e. leaf nodes) are favored.

For the biologically relevant regime, $2<\gamma<4$, trees have a symmetric structure. No single term in $\matW$ dominates. There is surprisingly little variation between the tree structures in this region. 

\begin{figure*}
  \centering
 \includegraphics[width = 0.6\textwidth]{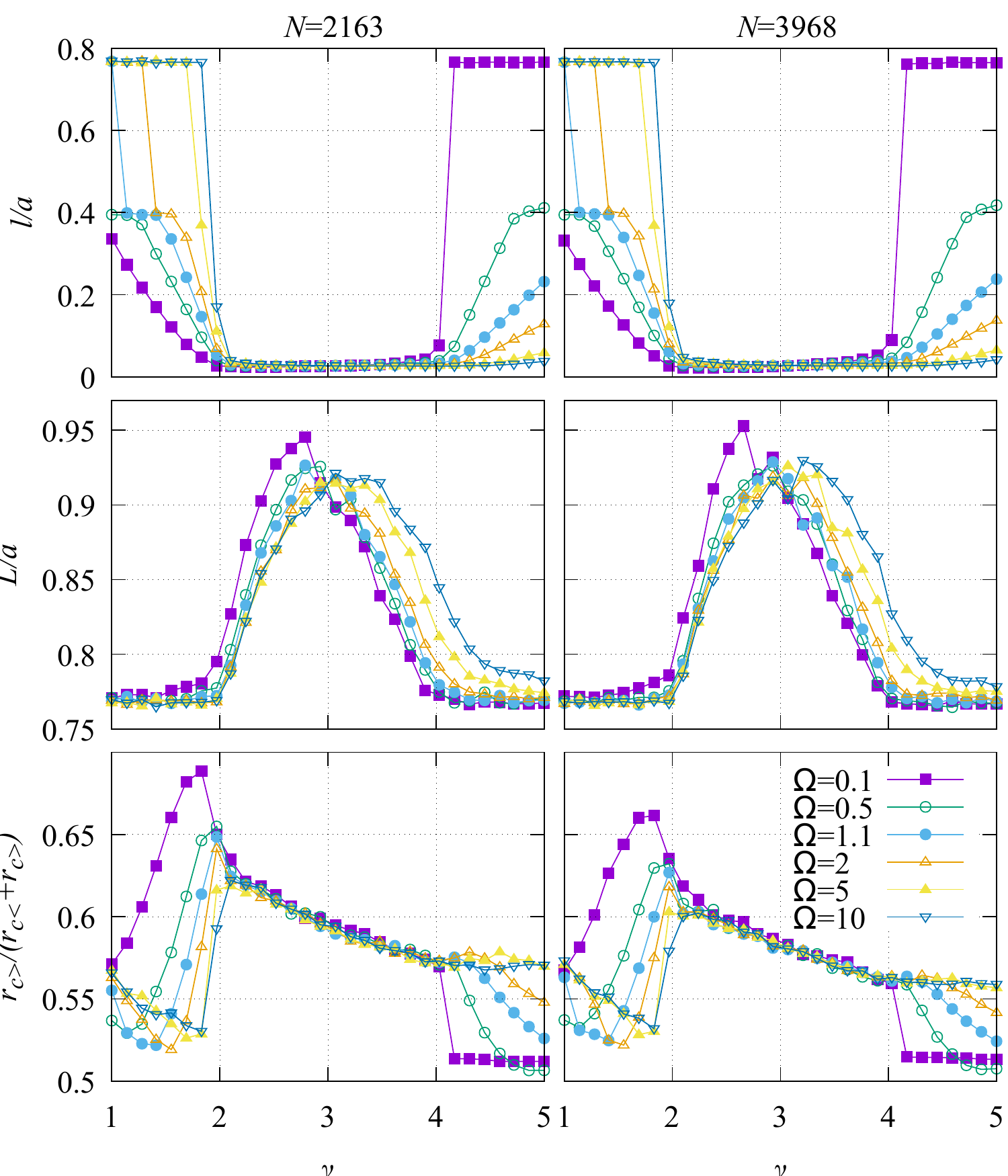}
  \caption{The tree morphology is highly sensitive to variation in the bifurcation exponent, but relatively insensitive to variation in $\Omega$ within the region of interest between $\gamma=2$ and $\gamma=4$. There is essentially no sensitivity to tree size.}
  \label{fig:bexp_sana}
\end{figure*}

To quantify the effect of varying $\gamma$ and $\Omega$ on the network structure, we have examined average segment length, path length and radius asymmetry. The radius of an arterial segment is given by $r_j$, and the length by $l_j$.  Average length is defined as $l=\sum l_j/N$. The average summed path length from root to leaf node is $L=\langle \sum_{\mathrm{path}} l_j \rangle $. Radius asymmetry is measured using $\langle r_{c>}/(r_{c<}+r_{c>})\rangle$ (where $r_{c>} \geq r_{c<}$). The sensitivities of these quantities to variations in $\gamma$ and $\Omega$ are presented in Figure \ref{fig:bexp_sana}.

In the typical range of biological tissue ($2<\gamma<4$), the dominant factor controlling morphological properties is $\gamma$. All morphological properties are insensitive to variation in $\Omega$. Average segment length is short and path length is long in this region, consistent with the branching structures seen for intermediate $\gamma$ in Fig. \ref{fig:varyntrees}. Bifurcation symmetry is in the range $0.58-0.62$, so bifurcations are moderately symmetric. Although $\Omega$ leads to minor changes in tree morphology in this regime, we note it can affect $\gamma_{\rm opt}$ and thus the tree morphology via $\gamma$ as a secondary effect.

In the regions $\gamma<2$ and $\gamma>4$, $\Omega$ is responsible for huge variations in the tree morphology, and $\gamma$ can also produce large variations in the various morphological and structural properties of the tree. Path length drops outside this region to approximately $a/\sqrt{2}$ consistent with a large number of straight paths from the root node to leaf nodes. For $\gamma<2, \Omega\ll 1$, the asymmetry increases dramatically. For all other regions of the parameter space, the asymmetry drops.

Morphological measurements are essentially insensitive to changes in $N$, consistent with additional segments adding more detail, but not qualitatively changing the tree structure. Panels on the left of Fig. \ref{fig:bexp_sana} show results for $N=2163$ and panels to the right for $N=3968$.

\begin{figure}
    \centering
    \includegraphics[width=80mm]{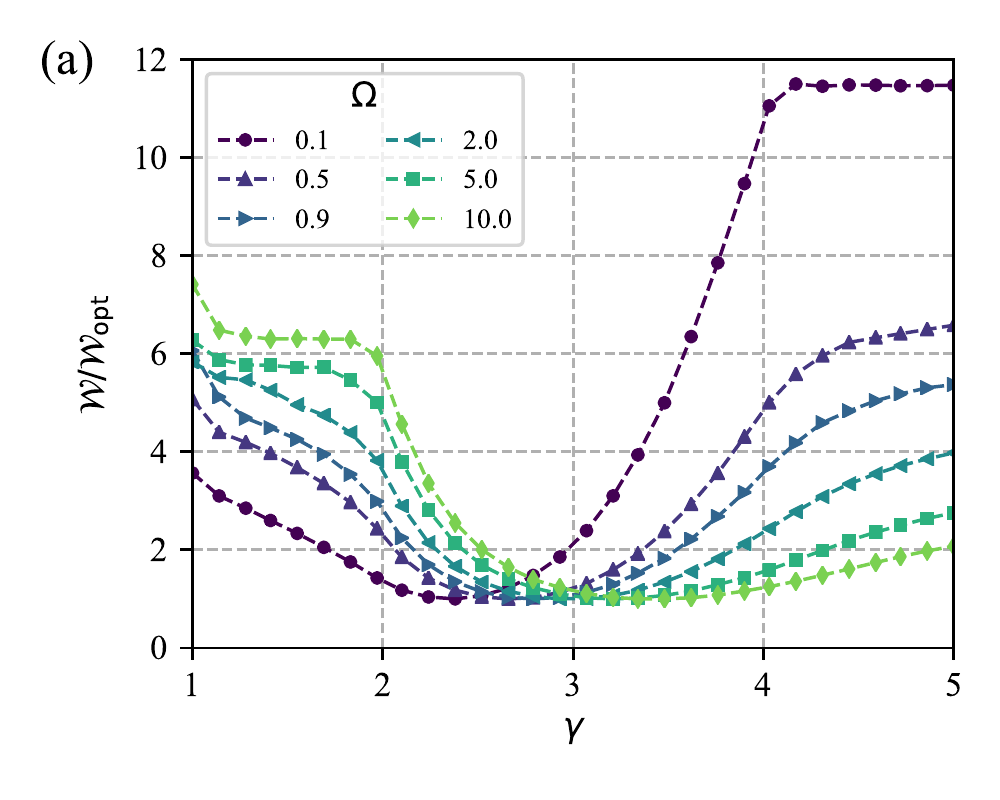}
    \includegraphics[width=80mm]{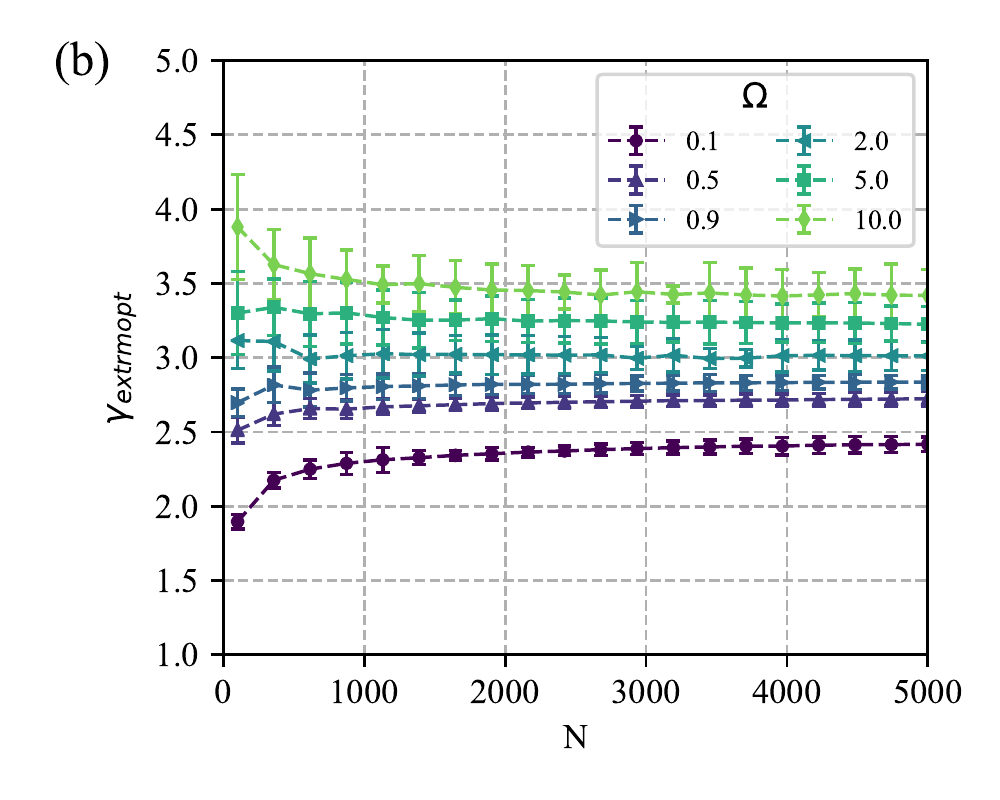}
    \caption{(a) A well defined global minimum in total power cost means that optimal bifurcation exponent $\gamma_{\rm opt}$ can be determined without ambiguity. The figure shows total power cost as a function of bifurcation exponent $\gamma$ for several values of $\Omega$. (b) The relationship of $\gamma_{\rm opt}$ to $N$ and $\Omega$, numerically determined using SALVO, is qualitatively similar to the relationship determined from analytic expressions. The figure shows  $\gamma_{\rm opt}$ vs $N$ for several $\Omega$.}
    \label{fig:opcurves_r}
        \label{fig:optgamc1c2}
\end{figure}

Optimal bifurcation exponent $\gamma_{\rm opt}$ can be determined without ambiguity from the minimum in $\matW$. Figure \ref{fig:opcurves_r}(a) shows how the total power cost varies with $\gamma$. There is a clearly defined global minimum for all values of $\Omega$ shown. $\gamma_{\rm opt}$ can be found by fitting a quadratic form to the bottom of the minimum.

The variation of $\gamma_{\rm opt}$ with $\Omega$ and $N$, numerically determined using SALVO, is qualitatively similar to the results from analytic expressions. Numerical values of $\gamma_{\rm opt}$ for various values of $\Omega$ vs $N$ are shown in Fig. \ref{fig:optgamc1c2}(b), and compare favorably to Fig. \ref{fig:asymmetricgammavsN}. Several numerical values are compared with the analytic results in Fig. \ref{fig:symmetricgamma}(a), also showing good agreement for both symmetric and asymmetric trees.

\begin{figure}
    \centering
    \includegraphics[height=50mm]{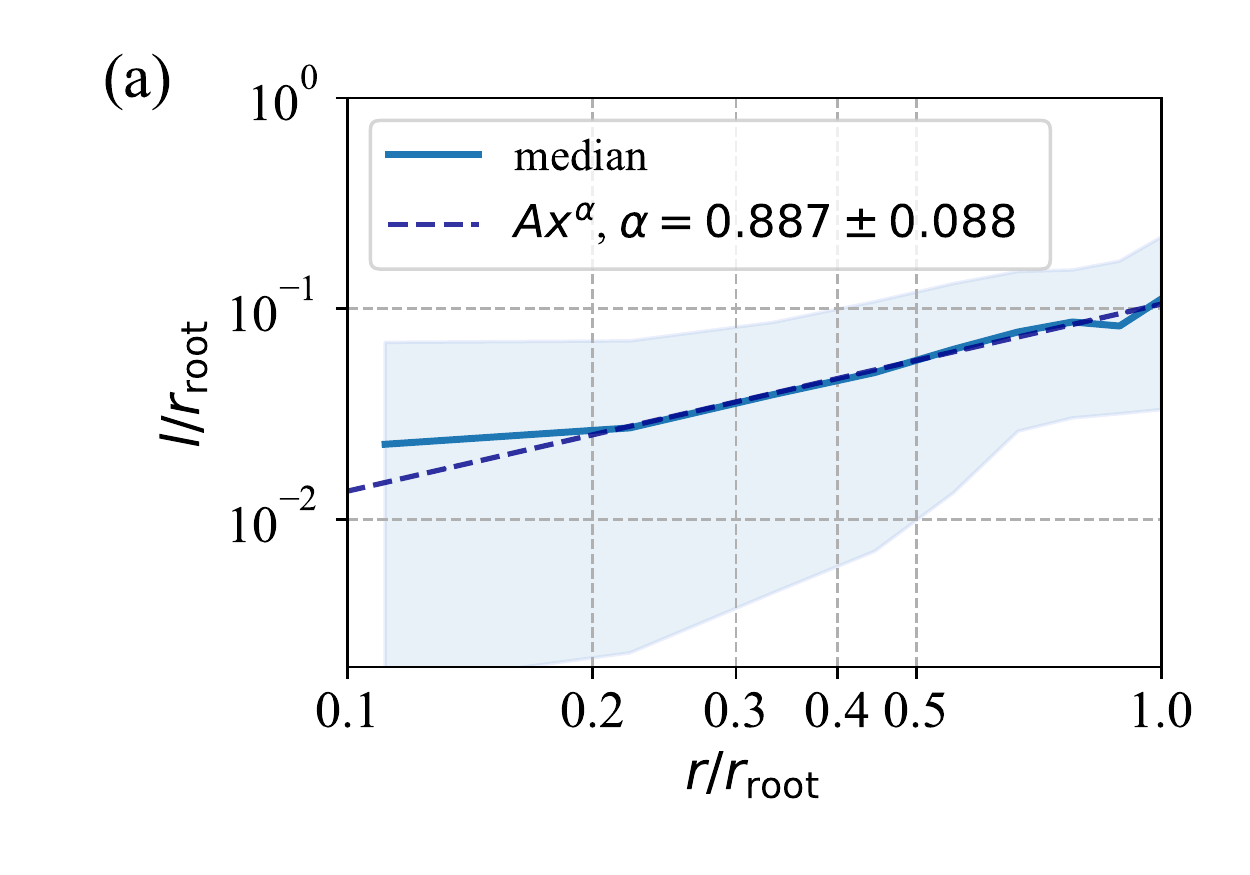}
        \includegraphics[height=50mm]{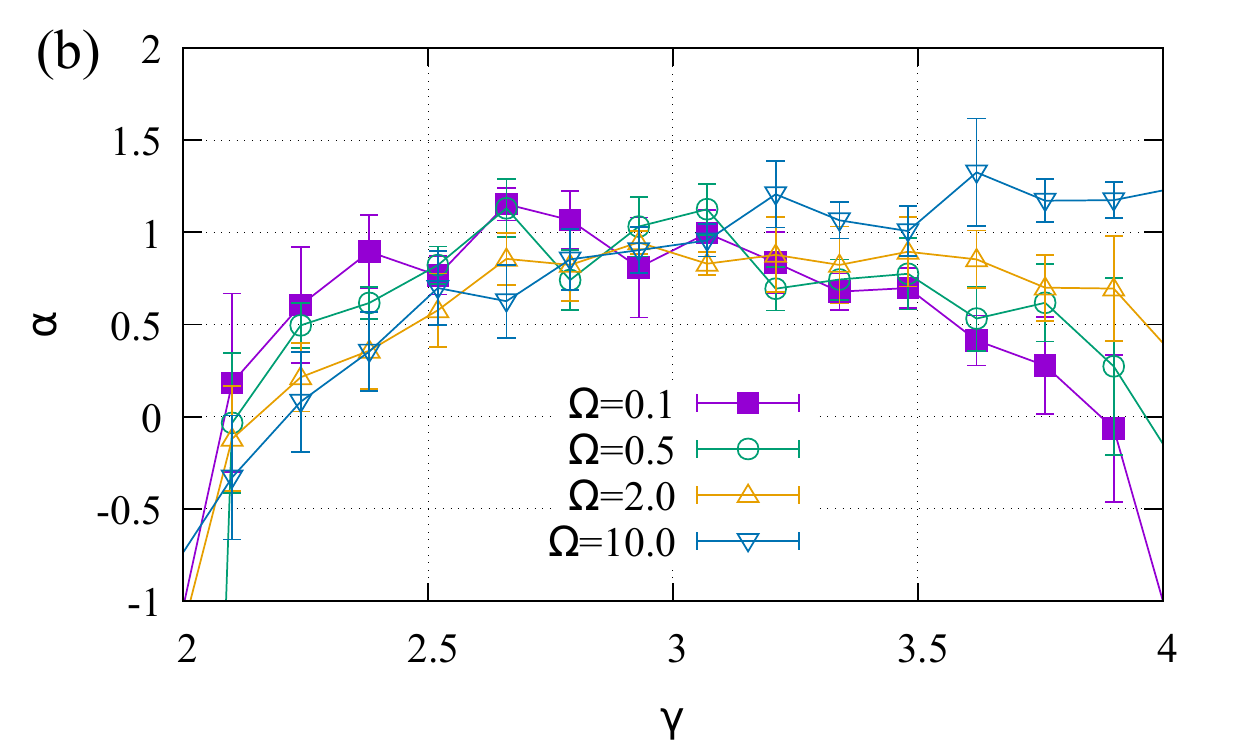}
    \caption{(a) A power law relationship is found for the median segment length in terms of segment radius calculated using SALVO. The figure shows median values of $l/r_{\rm root}$ vs $r/r_{\rm root}$, a power law fit (dashed line), and the 25th and 75th percentiles (light blue shading). To calculate the length--radius relation, segments are binned from trees with $N>2000$, $2.75<\gamma<3.25$, $\Omega=0.9$. (b) The length--radius exponent, $\alpha$, is close to one for trees grown with $2.5<\gamma<3.5$. To calculate the length--radius relation, segments are binned from trees with $N>2000$ and specific $\gamma$ and $\Omega$ values.}
    \label{fig:lvsrscatter}
    \label{fig:alphagrid}
\end{figure}

A power-law relationship, $l=Ar^{\alpha}$, is found for the median segment length in terms of segment radius calculated using SALVO (Fig. \ref{fig:lvsrscatter}(a)). The variation of the value of $l/r_{\rm root}$ with $r/r_{\rm root}$ is analyzed using Python 3. The expression $l=Ar^{\alpha}$ is fitted to the median value. Figure \ref{fig:lvsrscatter}(a) shows the fit. Regions shaded light blue show the range between the 25th and 75th percentiles in the length histograms. For segments selected from trees with $N>2000$, $2.75<\gamma<3.25$, $\Omega=0.9$, the fit has exponent $\alpha=0.887\pm 0.088$, consistent with experimental values \cite{Nakamura2014}. As in experiment, there is a strong scatter on the length.

The length--radius exponent, $\alpha$, is consistent with experimental values for trees grown with realistic $2.5<\gamma<3.5$, but can become effectively negative when long leaf segments dominate outside this region (Fig. \ref{fig:alphagrid}(b)). To calculate the length--radius relation, segments are binned from trees with $N>2000$ and specific $\gamma$ and $\Omega$ values. Where exponents are negative, the relation only poorly follows a power law, and errors on $\alpha$ are large. The relation is well followed within the region $2.5<\gamma<3.5$, and this leads to smaller error bars. Overall, errors on $\alpha$ are relatively large, and could be reduced by making calculations for additional $N$.

\section{Discussion and conclusions}


\fp{In this paper we determined analytic expressions, and carried out numerical calculations, for the properties and structures of globally optimal vascular trees, with the aim of understanding how overall complexity and physiological boundary conditions contribute to the optimal junction exponent and other structural properties of arterial trees.} Analytic expressions were derived for the special cases of maximally symmetric and asymmetric arterial trees.  The parameter space of the arterial trees was more fully explored by making numerical calculations with SALVO, enabling globally optimal vasculatures to be found for arbitrary tree morphology. Tree structures, morphological properties, sensitivity to dimensionless parameters and optimal bifurcation (junction) exponent are calculated.

\fp{Our analytic expressions are consistent with numerical calculations, and predict that $\gamma_{\rm opt}$ is insensitive to tree symmetry, so we propose that the analytic expressions derived here are applicable to a wide range of vasculatures.} Analytic expressions can be used for much larger trees, and would, therefore, be useful for predicting the properties of vasculatures within a range of organs where the number of vessel segments and overall complexity exceed the capabilities of current computers. We expect that it will be possible to extend the analytic expressions to include pulsatile flow and turbulence, and will investigate this possibility in future studies.

\fp{We predict that tree complexity is a significant contributor to the bifurcation exponents in living organisms.} The deviations we find from complexity are of similar size to those predicted by including turbulence and pulsatile flow in previous analyses. These deviations are particularly significant if physiological boundary conditions lead to $\Omega\neq 1$. This may occur since all organs, with their dramatically varying demands, are connected to the same major vasculature.  We expect that large variations of $\gamma_{\rm opt}$ with increasing complexity will also occur if a more detailed analysis including pulsatile flow and turbulence is carried out. 

\fp{We predict that arterial tree complexity can lead to optimal bifurcation exponent, $\gamma_{\rm opt} > 3$, a situation which can be found in experiment, and is of interest since inclusion of turbulence and pulsatile flow in single artery analyses leads to $\gamma_{\rm opt}<3$.} Large values of $\gamma$ are measured in e.g. the brain vasculature ($\gamma = 3.2$) \cite{keelan2019}, retina ($\gamma=3.1$ \cite{habib2006},  $\gamma=3.9\pm 0.12$ \cite{aldiri2010}) and other mammalian vasculatures where $\gamma$ can range as high as 4 \cite{Nakamura2014}. Such large $\gamma$ are not predicted by single segment analyses including effects related to pulsatile flow, elastic vessel walls and turbulence ($\gamma=2.3$), \cite{Nakamura2014}. Complexity and boundary conditions provide an additional contribution that can account for larger values of $\gamma_{\rm opt}$.

\fp{We predict that tree structures within the physiological regime are only sensitive to $\gamma$; outside the physiological regime structures are also highly sensitive to $\Omega$; and for all regimes tree structures are insensitive to $N$.} Changes in $N$ do not qualitatively change the morphology of the tree, but add more detail. Outside the regime $2<\gamma<4$, structure can change dramatically with $\Omega$.

\fp{Accurate values of $\gamma_{\rm opt}$ are particularly relevant to computational techniques used for growing very large arterial trees \emph{in-silico}, such as constrained constructive optimization (CCO).} Such algorithms rely upon a fixed bifurcation exponent to set the radii in the generated trees \cite{Schreiner1993,Schreiner2006,Karch2000}. Similarly, allometric scaling arguments require knowledge of $\gamma$ \cite{west1997general}, and variations of $\gamma_{\rm opt}$ could modify such approaches. $\gamma_{\rm opt}$ is quite hard to measure experimentally, leading to values with large uncertainties, and we consider the calculation of such values to be a useful application of our technique.

\fp{For values of $\gamma$ consistent with living systems, we find power law exponents in our computational trees that are consistent with the value $\alpha\sim 1$ obtained experimentally.} Experimental values range from $0.85<\alpha<1.21$ \cite{Nakamura2014}. We find a similar range of values in our numerical calculations, and with improved description of the flow, the accuracy of the predictions could be improved. Values of $\alpha$ are also useful as input to other calculations.

\fp{Future work to include additional physics, such as pulsatile flow, turbulence and vessel elasticity, would lead to a computational model with enhanced predictive power.} These improvements to the treatment of flow through vessels could be incorporated into both the analytic expressions derived in this paper, and into the cost function of SALVO without having to change the core algorithm. Once analytical expressions are modified to include this additional physics, we suggest that parameters such as $m_{b}$ could be determined from empirical results.

\fp{The significant structural changes visible at $\gamma\sim 2$ and $\gamma\sim 4$ would also be interesting areas for further study, since the rapid changes in the tree morphology are reminiscent of a phase transition.} These changes are on the edge of the physiologically relevant regime. Confirmation of a phase transition would require the identification of the order parameter and the signatures of critical behavior.

\fp{Finally, we hypothesize that evolutionary compromises may favor closer adherence to the predictions of single segment analyses in organs with large flow demands to the detriment of less flow-hungry organs.} Additional studies could be carried out to test this hypothesis. Overall, the computational and analytical approaches introduced here lead to a range of predictions regarding the structures of vascular trees, that provide interesting links to experimental and theoretical approaches.

\section*{Contributions}

JPH carried out the analytical calculations and led the study. JK carried out the numerical calculations. Both authors contributed to writing of the manuscript and analysis of the data.

\section*{Data availability}

The datasets generated and analyzed during the current study are available in the ORDO repository: \\
doi.org/10.21954/ou.rd.12220490 (note, these will be added at proof stage, data available on request).

\section*{Acknowledgments}

The authors have no competing interests. JK would like to acknowledge EPSRC grant no. EP/P505046/1.

\bibliography{references}

\end{document}